%% file: cscw15_venerandi.tex
\documentclass{sigchi}

\usepackage{balance}  
\usepackage{graphics} 
\usepackage{times}    
\usepackage{url}      
\usepackage{multirow}

\newcommand{\mytablesize}{\scriptsize}

\makeatletter
\def\url@leostyle{%
  \@ifundefined{selectfont}{\def\UrlFont{\sf}}{\def\UrlFont{\small\bf\ttfamily}}}
\makeatother
\def\pprw{8.5in}
\def\pprh{11in}

\usepackage[pdftex]{hyperref}
\hypersetup{
pdftitle={SIGCHI Conference Proceedings Format},
pdfauthor={LaTeX},
pdfkeywords={SIGCHI, proceedings, archival format},
bookmarksnumbered,
pdfstartview={FitH},
colorlinks,
citecolor=black,
filecolor=black,
linkcolor=black,
urlcolor=black,
breaklinks=true,
}

\toappear{\scriptsize Permission to make digital or hard copies of all or part of this work for personal or classroom use is granted without fee provided that copies are not made or distributed for profit or commercial advantage and that copies bear this notice and the full citation on the first page. Copyrights for components of this work owned by others than ACM must be honored. Abstracting with credit is permitted. To copy otherwise, or republish, to post on servers or to redistribute to lists, requires prior specific permission and/or a fee. Request permissions from Permissions@acm.org.  \\
CSCW'15, March 14 - 18 2015, Vancouver, BC, Canada. \\
Copyright 2015 ACM 978-1-4503-2922-4/15/03\ldots\$15.00. \\
http://dx.doi.org/10.1145/2675133.2675233 }
\begin{document}

\title{Measuring Urban Deprivation from  User Generated Content}

\numberofauthors{5}
\author{
   \alignauthor Alessandro Venerandi\\
       \affaddr{University College London}\\
       \affaddr{Gower Street}\\
       \affaddr{London WC1E 6BT, UK}\\
       \email{alessandro.venerandi.12@ucl.ac.uk}\\
   \alignauthor Giovanni Quattrone\\
       \affaddr{University College London}\\
       \affaddr{Gower Street}\\
       \affaddr{London WC1E 6BT, UK}\\
       \email{g.quattrone@cs.ucl.ac.uk}\\
   \alignauthor Licia Capra\\
       \affaddr{University College London}\\
       \affaddr{Gower Street}\\
       \affaddr{London WC1E 6BT, UK}\\
       \email{l.capra@cs.ucl.ac.uk}
   \alignauthor Daniele Quercia\\
       \affaddr{Yahoo Labs}\\
       \affaddr{Barcelona, Spain}\\
       \email{dquercia@acm.org}
   \alignauthor Diego Saez-Trumper\\
       \affaddr{Yahoo Labs}\\
       \affaddr{Barcelona, Spain}\\
       \email{dsaez-trumper@acm.org}
}


\maketitle

\section{Abstract}

\input{abstract}


~\newpage



\section{Introduction}
\input{introduction}

\section{Related Work}
\input{related}

\section{Datasets}
\input{dataset}

\section{Method}
\input{methods}

\section{Results}
\input{results}

\section{Implications, Limitations and Future Work}
\input{implications}


%
\section*{Acknowledgments}
This work has been carried out thanks to the Engineering and Physical Sciences Research Council (EPSRC) and the Centre for Urban Sustainability and Resilience (USAR) at University College London.

\balance
\bibliographystyle{acm-sigchi}
\bibliography{references}

\end{document}

%% file: abstract.tex
Measuring socioeconomic deprivation of cities in an accurate and timely fashion has become a priority for governments around the world, as the massive urbanization process we are witnessing is causing high levels of inequalities which require intervention. Traditionally, deprivation indexes have been derived from census data, which is however very expensive to obtain, and thus acquired only every few years. Alternative computational methods have been proposed in recent years to automatically extract proxies of deprivation at a fine spatio-temporal level of granularity; however, they usually require access to datasets (e.g., call details records) that are not publicly available to governments and agencies. 
To remedy this, we propose a new method to automatically mine deprivation at a fine level of spatio-temporal granularity that only requires access to freely available user-generated content. More precisely, the method needs access to datasets describing what urban elements are present in the physical environment; examples of such datasets are Foursquare and OpenStreetMap. Using these datasets, we quantitatively describe neighborhoods by means of a metric, called {\em Offering Advantage}, that reflects which urban elements are distinctive features of each neighborhood. We then use that metric to {\em (i)} build accurate classifiers of urban deprivation  and {\em (ii)} interpret the outcomes through thematic analysis. We apply the method to three UK urban areas of different scale and elaborate on the results in terms of  precision and recall.

%% file: introduction.tex
The world is undergoing a process of fast urbanization and it is estimated that by 2050 6.2 billion people will live in cities (68\% of the total global population) \cite{pdun11}. Although this process is supported by governments as it is expected to bring important advantages (e.g., better and less expansive public services, better living standards due to the concentration of economic activities) \cite{odi08}, recent research has also shown that inequality is dangerously on the rise, with some areas benefiting substantially more than others from public investments and economic growth \cite{unhsp08}. Quantifying urban poverty promptly, and at a fine level of spatial granularity, has thus become a priority for governments worldwide, so to be able to monitor the impact of urbanization, and to make data-driven decisions as to how to allocate limited financial resources for regeneration projects. 

Traditionally, socioeconomic deprivation has been measured using data acquired through household surveys; while such data is semantically rich, it is also very expensive to obtain and process. As a result, it is acquired with a rather low  frequency that varies from few years for developed countries like the UK, to several years for developing countries like Cote d'Ivoire. To remedy this, computational social scientists have started to develop new methods that aim to automatically derive metrics of deprivation from alternative data sources, that afford finer spatio-temporal granularity than survey data. Data sources used to date span from call detail records (e.g., \cite{eagle10,mao13,smith13a}), to satellite images (e.g., \cite{elvidge97,elvidge01,noor08}), to transit data (e.g., \cite{smith13b}). A common limitation to all these methods is the reliance on datasets that are very difficult to obtain, thus severely limiting the applicability of the methods themselves. 

In this paper, we propose a new method that aims at accurately computing urban deprivation at a fine-grained spatio-temporal granularity, while relying on easily accessible datasets. More precisely, in terms of datasets, our method relies on user-generated content that captures which urban features are present in a neighborhood. Examples of such datasets are Foursquare and OpenStreetMap. We made this choice inspired by qualitative works in the public health domain that have found important relationships between the presence of certain urban elements in a given area, and the socioeconomic well-being of its residents. For example, researchers have found that health-promoting amenities (e.g., golf courses in Australia \cite{giles02}, fitness centers and dance facilities in USA \cite{powell06}) are more concentrated in well-off areas; on the contrary,  potentially health-harmful resources (e.g., fast food outlets in England and Wales \cite{cummins05}) are more concentrated in poorer areas. From these datasets, our method automatically extracts a  metric called {\em Offering Advantage} that intuitively reflects which urban elements are distinctive features of each neighborhood. Using correlation analysis, we prune down these features so to consider only those that are significantly correlated with urban deprivation; we then use those significant features to build classifiers of urban deprivation and finally, by means of thematic analysis,  interpret the outcomes. We illustrate how to apply the method in practice in three UK urban areas of different scale  (i.e., Greater London, Greater Manchester, West Midlands); finally, for these three case studies, we elaborate on the precision and recall of the results.

In the reminder of the paper, we first provide an overview of related works in this domain. We describe the datasets, and the method developed to leverage them. We then present the results of our evaluation, before concluding the paper with a discussion of implications, limitations and future work.

%% file: related.tex
In an attempt to measure socioeconomic deprivation at a fine level of spatio-temporal granularity, researchers have started to develop computational methods that automatically mine a variety of data sources, looking for significant and strong signals of deprivation. To this end, three main sources of data have been used.

{\em Call Detail Records (CDRs).} These are records about calls and text messages measured by telecommunication providers for billing purposes. CDRs contain information about time, duration, caller ID, callee ID and location of the antenna tower through which the call or the text message is sent. Features extracted from these datasets have then been used to assess socioeconomic well-being of populations. A pioneering work in this domain has been conducted by Eagle \emph{et al.}~\cite{eagle10};  they studied the relationship between CDRs from land lines and mobile phones in England and the Index of Multiple Deprivation (IMD). Their results highlight a strong correlation between call network diversity and deprivation, confirming the hypothesis that having a varied set of contacts is a signal of socioeconomic well-being. A few years later, Mao  \emph{et al.} investigated the relationship between features of calls in Cote d'Ivoire and some economic indexes of ten areas of high economic activity~\cite{mao13}. They discovered that the ratio of  outgoing calls per area, relative to the total of  outgoing plus  incoming calls, has high correlation with annual income. Smith  \emph{et al.} analyzed the same mobile phone dataset and found that features of network diversity and introversion (i.e., ratio of within-area calls vs. inter-area calls) strongly correlate with deprivation~\cite{smith13a}. A common limitation to these studies is the difficulty to gain wide access to CDR data, as telecommunication operators do not tend to make that data publicly available. 

{\em Satellite imagery.} A completely different approach has looked into analyzing patterns from satellite images, in order to  map economic development. In particular, researchers have extracted a feature called Night Time Light (NTL) from images,  that is, the total surface lit during night time. Elvidge  \emph{et al.} found a correlation between  NTL and countries' Gross Domestic Product~\cite{elvidge97, elvidge01}. Similarly, Noor  \emph{et al.} studied the relationship between NTL and a composite index of wealth for several administrative regions of some African countries \cite{noor08}. The correlations found were initially high; however, more recent research showed far lower correlations. These findings suggest that, as the penetration of electrical lightning reaches saturation, the signal present in these datasets disappears. 
Satellite images are as hard to get as CDRs; furthermore, the methods seem only applicable in under-developed countries up to the point where electricity becomes a commodity.

{\em Transit data.} Another source of information which researchers have been analyzing to get insights into deprivation is urban transit data. Smith  \emph{et al.}~\cite{smith13b} used Oyster Card data  (i.e., electronic ticketing system capturing journeys made within the London public transport network) to derive features of mobility flow between areas, and of transport modality choice. They then used these features to build a classification model to identify highly deprived areas, as measured by the UK Index of Multiple Deprivation.  Although the model achieves high prediction accuracy, it can only estimate deprivation for 10\% of London (i.e., where a tube station is present). Furthermore, similar to works on CDRs data, this line of work requires access to datasets that are very difficult to get hold of (if available at all).

All lines of research above require access to datasets that are very difficult to get. An alternative approach is to rely on more easily accessible user-generated content (UGC) for the same purpose. User-generated content comes in a wide variety of forms, thanks to the big popularity and uptake of social media applications, especially in developed countries. These data sources have been used by researchers to develop a better understanding of our cities. For example, geo-coded tweets have been analyzed to  quantify sentiment/mood, as it varies between  neighborhoods of different socioeconomic standing~\cite{quercia12a}; Twitter has also been analyzed to investigate the relationship between the topic of discussion in a certain area, and the deprivation score of that area~\cite{quercia12b}; Foursquare check-ins have been used to redefine neighborhood boundaries, by categorizing different areas of a city through a clustering model that leverages similarity of urban functions, rather than using super-imposed administrative boundaries~\cite{cranshaw12}. 

In line with this last stream of research, we propose to use user-generated content to mine urban deprivation. Inspired by previous qualitative works in the public health domain  that found correlations between the presence of certain urban venues and deprivation~\cite{giles02, powell06}, we propose to automatically extract such urban features directly  from easily accessible user-generated content datasets, specifically,  Foursquare and OpenStreetMap. In the next section, we describe the datasets, before providing the details of our method.

%% file: dataset.tex
To conduct this work, we needed access to two types of datasets: on one hand, indicators of socioeconomic deprivation at a fine-grained spatial granularity; on the other hand, detailed records of which physical elements are present in  the built environment. We use the Index of Multiple Deprivation for the former, and both Foursquare and OpenStreetMap for the latter.

\subsection{Index of Multiple Deprivation (IMD)}

As measure of deprivation, we use the UK Index of Multiple Deprivation (IMD),\footnote{\url{https://www.gov.uk/government/uploads/system/uploads/attachment_data/file/6871/1871208.pdf}} computed at the level of small census areas known as Lower-layer Super Output Areas (LSOAs). LSOAs were defined to roughly include always the same number of inhabitants (around 1.500)~\cite{mclennan2011english}. IMD is a composite score, calculated as the weighted means of seven distinct domains: income deprivation, employment deprivation, health deprivation, education deprivation, barrier to housing and services, crime, and living environment deprivation. The higher the IMD score, the more deprived the neighborhood, and viceversa; overall, IMD scores follows a normal distribution~\cite{mclennan2011english}. For the purpose of our study, we collected IMD scores for three UK urban areas that differ in terms of population size and geographic span, so to test the applicability of our approach to case studies of different scales. Those are: Greater London, 
Greater Manchester, 
and West Midlands. 

We chose Greater London as an example of large metropolitan city. Greater Manchester and West Midlands are both examples of mid-size cities instead, albeit with a rather different population density. General information about these cities is provided in Table~\ref{tab:info_cities}.\footnote{\url{http://www.ons.gov.uk/ons/dcp171778_270487.pdf}} 

\begin{table}[!ht]
	\centering
	\mytablesize
    \begin{tabular}{l|ccc}
        {\em Urban area} & {\em Population} & {\em Density} & {\em Area}\\
        \hline
        Greater London     &  8,204,100    & 5,218 ppl per km$^2$ & 1,572 km$^2$  \\
        Greater Manchester &  2,685,400    & 2,105 ppl per km$^2$ & 1,276 km$^2$ \\
        West Midlands      &  2,738,100    & 3,039 ppl per km$^2$ & 902 km$^2$ \\
    \end{tabular}
    \caption{Population, population density, and area for the three UK urban areas.}	
    \label{tab:info_cities}
\end{table}

\subsection{Foursquare}

Foursquare is a mobile social-networking application launched in 2009 and is also one of the most popular location-based social-networking websites.\footnote{\url{https://foursquare.com/about}} In Foursquare, when registered users visit a venue, they can `check-in' on the mobile application to share their location with their friends. In April 2012, Foursquare reported 20 million registered users, with more than 2 billion check-ins~\cite{Lacy2011}. Aside from checking-in at existing venues, Foursquare users can also create new ones. Possible conflicts in the definition of venues are solved in a bottom-up fashion: the more accurate a  description, the more likely users will be able to recognize (check-in to) it. Foursquare then attempts to merge multiple descriptions that are likely to refer to the same venue using a Venue Harmonization procedure,\footnote{\url{ https://developer.foursquare.com/overview/mapping}} which includes the use of developer-contributed geographic databases. At the minimum level, each venue needs to be defined by a pair of latitude/longitude coordinates, a name, and a category (e.g., Church, School, Pub). Janne Lindqvist \emph{et al.} have recently studied why people check-in and found five factors, one of which is particularly relevant to the creation of places~\cite{lindqvist2011m}: individuals tend to use Foursquare to see where they have been in the past and ultimately curate their own location history. In cities where Foursquare has high penetration, the venues recorded in this service should thus collectively form a  well-curated land use dataset. For the purpose of this work, we use the official Foursquare API to crawl all Foursquare venues for the three UK urban areas under consideration.\footnote{\url{https://api.foursquare.com}} We performed this step between 04/03/2014 and 08/04/2014; a summary of the dataset obtained is reported in  Table~\ref{tab:fsq}. Given that the three cities show different orders of magnitude in the number of venues, applying our method to them is likely to  translate into interesting  insights about our method's applicability to different urban contexts.

\begin{table}[!ht]
	\centering
	\mytablesize
    \begin{tabular}{l|ccc}
        {\em Urban area} & {\em \# Venues} & {\em \# Check-ins} & {\em \# Categories}\\
        \hline
        Greater London     &  178.756      & 26.344.132   & 503  \\
        Greater Manchester &   43.874      & 3.235.174    & 421  \\
        West Midlands      &   37.370      & 2.424.546    & 435  \\
    \end{tabular}
    \caption{Number of Foursquare venues, number of check-ins, and number of Foursquare categories across the three UK urban areas.}	
    \label{tab:fsq}
\end{table}



\subsection{OpenStreetMap}

OpenStreetMap (OSM) is perhaps one of the most successful examples of geographic crowd-sourcing, with currently over 1.6M users, collectively building a free, openly accessible, editable map of the world.\footnote{\url{http://download.geofabrik.de/}}  OSM data covers three types of spatial objects:  {\em nodes},  {\em ways}, and {\em relations}. Nodes broadly refer to Points of Interests (POIs), ways are representative of roads, and relations are used to group together other objects (e.g., administrative boundaries, bus routes). For the purpose of this study, only nodes are relevant. Similar to Foursquare venues, an OSM node consists of three main attributes: a geographical position (latitude and longitude), a name, and a category (called amenity type in OSM jargon). Differently from Foursquare, OSM categories are not chosen through a given taxonomy, and contributors are free to use whatever words they find most suitable. We downloaded OSM node data for the three cities under consideration on 07/05/2014; summary statistics are provided in Table~\ref{tab:osm_users_poi_edits}.

\begin{table}[!ht]
	\centering
	\mytablesize
    \begin{tabular}{l|cc}
        {\em Urban area} & {\em \# Nodes} & {\em \# Categories} \\
        \hline
        Greater London     &  79.343       &    896 \\
        Greater Manchester &  24.321       &    381 \\
        West Midlands      &  27.885       &    465 \\
    \end{tabular}
    \caption{Number of OSM nodes and number of OSM categories across the three UK urban areas.}	
    \label{tab:osm_users_poi_edits}
\end{table}



\begin{figure}[t]
    \centering
    \includegraphics[width=0.45\columnwidth]{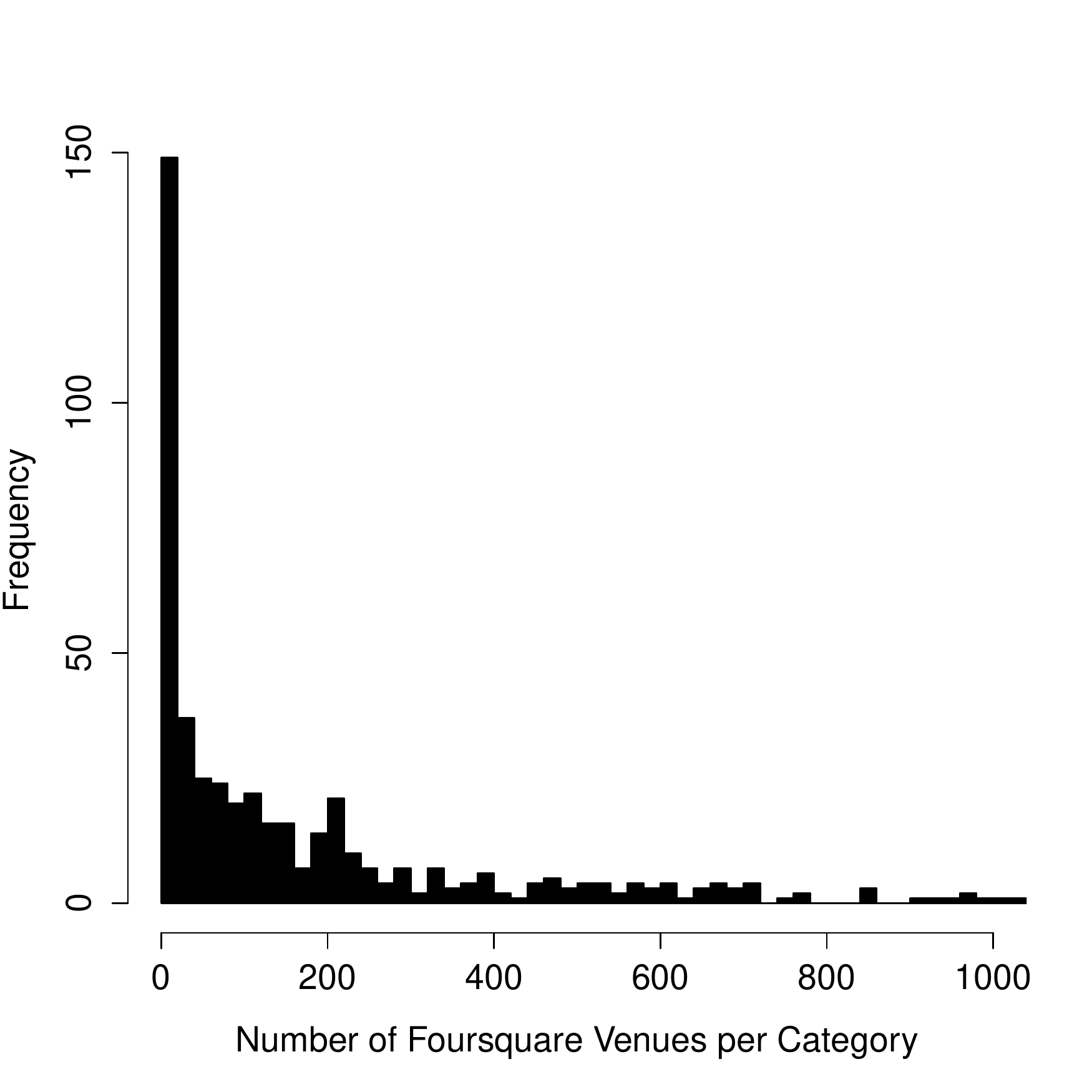}
    \includegraphics[width=0.45\columnwidth]{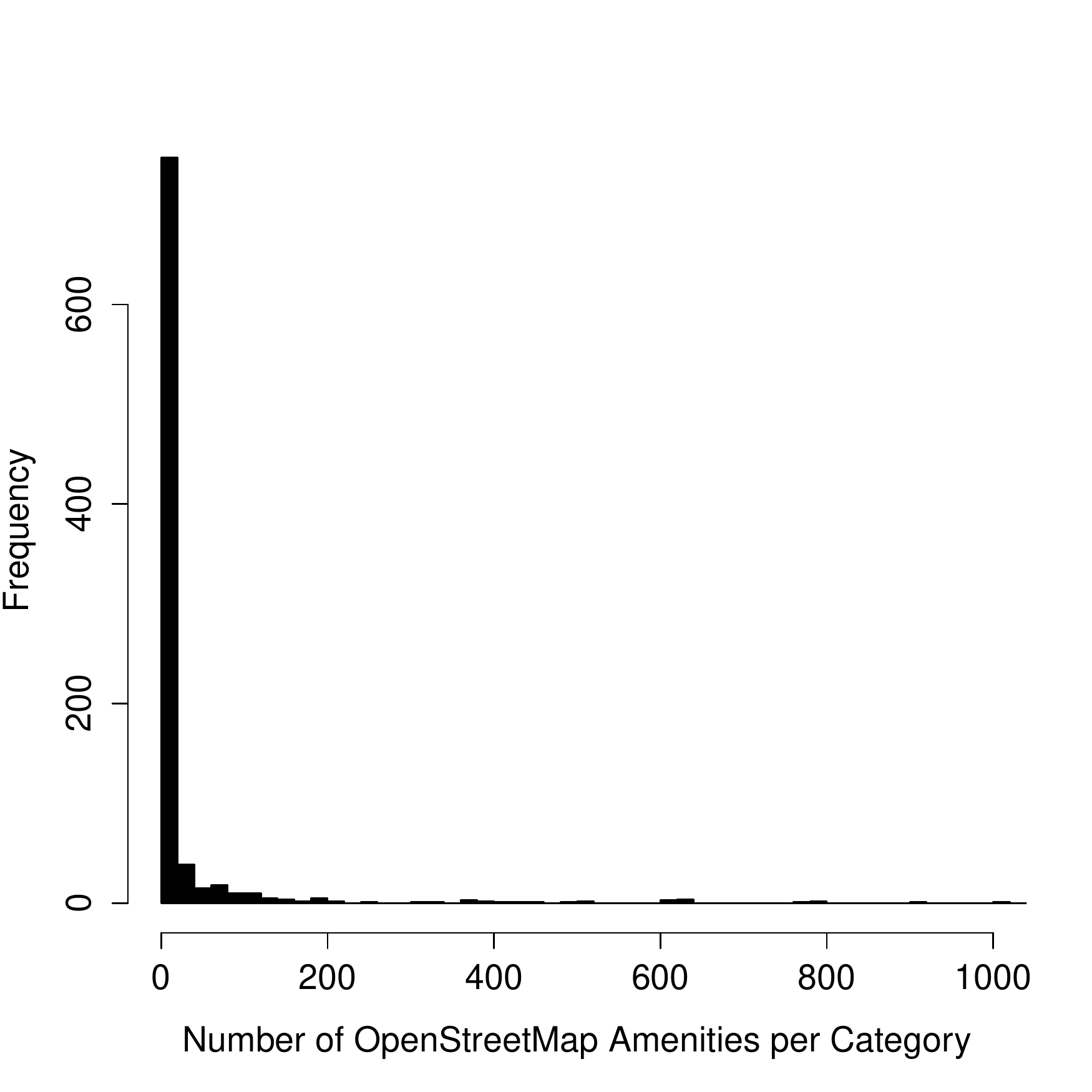}
    \caption{Frequency distribution of Foursquare categories (left) and OSM categories (right).}
    \label{fig:amenityDist} 
\end{figure}

Preliminary analysis on Foursquare and OSM datasets for Greater London (Figure~\ref{fig:amenityDist}) shows that categories follow a long-tailed distribution, with few categories having large number of venues/nodes, and many categories with very few venues/nodes instead. Furthermore, we notice that, although Foursquare and OSM conceptually provide the same kind of information (e.g., physical elements present in the urban environment), their users map different things. In fact, while the most frequent Foursquare categories are restaurants, pubs and cafes, the most frequent OSM categories are bus stops, crossings and post boxes. This is not surprising, given the fundamentally different purposes behind these services: Foursquare is mainly used to share locations with which users like to be associated; conversely, OSM is mainly used to capture in full what is present in the world. Our work takes advantage from the complimentary nature of these two UGC datasets, as we hypothesize that more accurate area profilers can be built by combining the peculiarities of the two datasets, rather than using them in isolation, thus potentially unveiling a broader set of features that correlate with urban deprivation.

%% file: methods.tex
In this section, we describe the method we have developed to estimate urban deprivation from UGC data (Foursquare and OSM in particular). We begin with a description of the adopted spatial unit of analysis, and of the motivation behind this choice.  We then define the metric implemented to automatically extract, from UGC data, the urban features characterizing these spatial units. Finally, we provide a step-by-step description of how to apply our method in practice: from  correlation analysis to identify urban features related to deprivation, to  classifiers to estimate deprivation levels, to  thematic analysis to interpret  results.

\subsection{Unit of Analysis}

The goal of this work is to measure the socioeconomic deprivation of different areas within a city by capturing the urban elements that are physically present in each neighborhood.  To do so, we need to define a spatial unit of analysis that is representative of a city neighborhood. As mentioned in the previous section, IMD is available at a fine-grained spatial granularity, that of  LSOAs. However, these units, which have been introduced relatively recently (in 2001), are too small and too arbitrarily defined to be meaningful to urban residents. Indeed, we calculated the average LSOA for Greater London to be no bigger than 33 hectares; furthermore, they are identified through alphanumeric codes (e.g., E01008881) which do not represent neighborhoods as recognized by citizens. We therefore choose a different spatial unit called {\em ward}. Wards have a much longer history compared to LSOAs (they have existed since the Middle Ages).  Wards are defined by the UK Government\footnote{\url{http://www.ons.gov.uk/ons/guide-method/geography/beginner-s-guide/administrative/england/electoral-wards-divisions/index.html}} and represent both electoral subdivisions and ceremonial entities; their geographic extension exceeds that of the LSOAs with an average area size for Greater London of about 250 hectares. Finally, wards are identified through human-understandable toponyms (e.g., Highgate). For all these reasons, we argue that wards, more than LSOAs, are good representations of citizens' neighborhoods. Using official geographic definitions of wards in the UK,\footnote{\url{https://geoportal.statistics.gov.uk/Docs/Boundaries/Wards_(E+W)_2011_Boundaries_(Full_Extent).zip}} we computed 625 wards for Greater London, 215 for Greater Manchester and 163 for West Midlands. For simplicity, in the next sections we will refer to wards as `neighborhoods'.


\subsection{Offering Advantage Metric}

Having defined our spatial unit of analysis, we now need to profile each ward in terms of the urban features that characterize it. In doing so, we aim to capture not just what urban elements are physically present in a neighborhood, but more importantly what elements make it distinct with respect to other neighborhoods. As shown in the previous section, some categories are much more frequent than others, so that a simple count of what amenities are present is not sufficient to elicit distinctiveness. Rather, we propose to use a metric called {\em Offering Advantage}, which weights categories by their popularity; intuitively,  the presence of one element from an unpopular category is much more significant in profiling a neighborhood than the presence of one element from a very popular category. In practice, this measure relies on a concept used in economics called Revealed Comparative Advantage (RCA) \cite{hidalgo2007product}. This is used to measure whether a country exports more of good $i$ (as a share of its total exports), than the average country; if so, then $RCA > 1$. 

The RCA of a country $c$ is usually evaluated with this formulation:
	$$
		RCA_{c,i} = \frac{goods_{c,i}}{ goods_c } \cdot \frac{ world }{world_i}
	$$
where $goods_{c,i}$ denotes how many goods $i$ are exported by  country $c$; $goods_c$ denotes the total number of goods exported by the country $c$; $world$ is the total number of goods exported all around the world; finally, $world_i$ indicates how many goods $i$ are exported all around the world. 

In our context, this measure reflects to what extent a neighborhood $n_k$ provides more (Foursquare/OSM) POIs of a certain category $c_i$ than the average neighborhood. More specifically:
	$$
		OA(c_i,n_k) = \frac{count(c_i,n_k)}{ \sum^{N}_{j=1}{count(c_j,n_k)} } \cdot \frac{ \sum^{N}_{j=1}{count(c_j)} }{count(c_i)}
	$$
where $OA(c_i,n_k)$ denotes the {\em Offering Advantage} of a (Foursquare/OSM) POI category $c_i$ in the neighborhood $n_k$; $count(c_i,n_k)$  counts how may POIs of category $c_i$ are present in the neighborhood $n_k$; $N$ is the total number of (Foursquare/OSM) POI categories; finally, $count(c_i)$ counts how many (Foursquare/OSM) POIs of category $c_i$ are present in the whole urban area. This metric has been recently used in a preliminary study on a Foursquare dataset for Greater London \cite{quercia2014mining} and showed to provide better association with deprivation compared to a raw count of number of POIs for each category in a ward. We thus rely on the same metric to study deprivation and we apply it on two datasets of user generated content  and on three cities.

\subsection{Approach}

Having defined our spatial unit of analysis (i.e., UK wards) and the metric we use to quantitatively profile these areas (i.e., {\em Offering Advantage}), we can now describe our proposed approach. 

\subsubsection{Correlation Analysis to Identify Significant Urban Features} 

Each area (ward)  is described by a vector that reports the OA metric for each Foursquare/OSM POI category in that area. Although there are several hundred POI categories, not all of them will bear significant signals of deprivation. For example, the very same POI category (e.g., school) may be equally present in well-off and deprived wards. To filter out all those categories that do not consistently signal deprivation within the urban area under consideration, we use correlation analysis between the OA metric automatically derived from user-generated content, and the Index of Multiple Deprivation (IMD) that acts as ground truth in this context. To do so, the following three steps had to be performed: first,  we had to reconcile the spatial unit of analysis at which IMD is available (i.e., LSOA), with the spatial unit of analysis used in this work (i.e., ward). We did so by computing the IMD score of a ward as the average of the IMD scores of the LSOAs it spatially contains. One may wonder if, by doing so, we cause significant data loss and inaccuracies. We found this not to be the case, since deprivation scores for LSOAs which belong to the same ward are very consistent (the standard deviation of IMD values related to LSOAs contained within wards is smaller than the corresponding average value, for all wards). 

Second, since we are dealing with geographical data, we had to address the problem of spatial auto-correlation in our data. This, in fact, can lead to incorrect conclusions. Spatial auto-correlation is the tendency for measurements located close to each other to be correlated, a property that generally holds for variables observed across geographic spaces \cite{legendre93}. In broader terms, this is the direct quantitative demonstration of Tobler's First Law of Geography, which states that `everything is related to everything else, but near things are more related than distant things' \cite{tobler70}. When high spatial auto-correlation occurs, traditional metrics of correlation (e.g., Pearson and Spearman) that require independence in the observations cannot be applied. We tested our data for spatial auto-correlation and we found it to be indeed high. To overcome this problem, we used a method familiar to natural scientists \cite{grenyer06} and introduced by Clifford {\em et al.} \cite{clifford89}. This approach addresses the `redundant, or duplicated, information contained in geo-referenced data' \cite{griffith11} -- the effect of spatial auto-correlation -- through the calculation of a reduced effective sample size. The significance of the correlation coefficients presented in the next section is obtained through the implementation of Clifford {\em et al.}'s method, partly accounting for spatial auto-correlations. 

Third and lastly, since we are performing simultaneous correlation tests with multiple variables (i.e., our POI categories), the chance of  incorrectly rejecting the null hypothesis for some of these variables (and thus of obtaining false positive results), increases. To quantify this threat,  we implement a statistical control technique commonly used among researchers dealing with datasets comprising a large number of distinct variables (e.g., in genomics) called False Discovery Rate (FDR)~\cite{storey2003}. In practice, the method analyses the distribution of $p$-values of the tested variables, and produces a list of so-called $q$-values, each varying between 0 and 1,  indicating the expected proportion of false discoveries within the list of findings. In the next section, we will report $q$-values, computed using the FDR method, over the list of variables found to be significantly correlated when using the Clifford {\em et al.} correlation method.

Note that there exists a temporal discrepancy between the IMD dataset (2011), and the Foursquare/OSM datasets (2014). We expect this discrepancy to have limited impact on the findings we will report in the next section,  as IMD values did not significantly vary in the last two reporting periods (i.e., 2008 and 2011).


\subsubsection{Classification Tool of Deprivation} 

Once we have identified the POI categories that correlate with deprivation in the urban areas, we can then build a classifier of deprivation. One might wonder what the advantage of such a classifier is, if we still need IMD scores to be able to identify what urban features signal deprivation in the first place. We envision two uses of this approach that would make it more affordable than running citywide household surveys every (few) year(s): a first approach would require the completion of household surveys only in a small subset (e.g., 25\%) of the city neighborhoods; our method would then apply correlations analysis to derive features upon which to build classifiers for the remaining  (e.g., 75\%) areas. A second approach would see the completion of citywide household surveys, from which to compute IMD scores manually, only once every several years: at these times, correlation analysis would be conducted. Once this is done, for the several years in between surveys, the classifiers would then be used to estimate deprivation at no extra cost. This second approach is based on the assumption that, although the deprivation score of an area may vary year by year (e.g., as a result of processes of urbanization, migration, and gentrification), the relationship between deprivation scores and urban features is much more stable; that is, certain POI types remain concentrated where wealth -- or deprivation -- is higher (although we do not know whether that is because people move towards area with certain POI types, or because areas with some POI types  attract people of certain economic status, or both).

\subsubsection{Thematic Analysis to Derive Significant Themes} 

The correlation analysis potentially identifies tens of Foursquare/OSM POI categories that are associated with deprivation; some of these might be redundant (e.g., bus, bus stop), and some others might be due to chance. The outcome of the classification tool built atop of these categories  might thus be fragmented and difficult to interpret. To avoid this, our proposed method requires the application of inductive thematic analysis~\cite{bc06} to the filtered Foursquare/OSM POI categories, so to group together those categories that are semantically related. The result is a very small set of meta features, or themes, that represent simple and distinctive urban characters. 

In the next section, we apply our method to  three UK urban areas of different scale (i.e., Greater London, Greater Manchester, West Midlands), and discuss precision and recall of the results.

%% file: results.tex

We first illustrate the results of the correlation analysis that is carried out to identify which POI categories (out of the hundreds present in Foursquare and OSM) are significantly associated with deprivation. Second, we build classification tools based on the identified categories, and assess their accuracy compared to a baseline classifier and to a state-of-the-art approach. Finally we illustrate which meta-features emerged from the thematic analysis.

\subsection{Finding Signals of Deprivation}

We explore to what extent we can exploit land use data extracted from user-generated content to get useful insights about the socioeconomic status of city neighborhoods. To this end, we calculated {\em Offering Advantage} for all Foursquare and OSM categories for each urban area and, through the Spearman's rank correlation coefficient $r_s$, we correlated it with IMD. Tables~\ref{tab:numberOfCorrelations4sq} and~\ref{tab:numberOfCorrelationsOSM} show the number of Foursquare and OSM categories correlated (positively or negatively) with IMD, grouped by strength of correlation. From the analysis of these two tables, we highlight two important observations: first, there is a higher number of categories significantly correlated with deprivation in Foursquare than in OSM; this suggests that most OSM categories are conceptually less associated with socioeconomic aspects of cities. Second, the number of weak-to-moderately correlated categories ($r_s \in [0.2,0.4)$ or ($r_s \in [0.4,0.6)$) is high across all cities (15 categories for Greater London, 24 for Greater Manchester, and 34 for West Midlands), suggesting there is indeed a wealth of urban features we can mine from UGC to study deprivation for cities of different scales (in terms of area size, population, and user-generated content). Tables~\ref{tab:correlationsFoursquare} and~\ref{tab:correlationsOSM} show the top three categories most positively and most negatively correlated with IMD, using Foursquare and OSM data respectively. Note that, while Foursquare categories provide detailed information about services and facilities (e.g., Student Center, Caribbean), OSM categories tend to give more information about road system elements (e.g., traffic signals, crossing). The two datasets thus offer complimentary information at times, and they should thus be studied together.

\begin{table}[h]
 \mytablesize
 \tabcolsep 3pt
 \centering
 \begin{tabular}{c|ccc}
                    & $|r_s| \in [0.05,0.2)$ & $|r_s| \in [0.2,0.4)$ & $|r_s| \in [0.4,0.6)$\\
 \hline
 Greater London     & 23                  & 15                  & 0 \\
 Greater Manchester & 30                  & 24                  & 0 \\
 West Midlands      & 17                  & 33                  & 1 \\
 \end{tabular}
 \caption{Number of Foursquare POI categories (positively or negatively) correlated with IMD (all results shown are statistically significant, $p$ $<$ 0.05)}
 \label{tab:numberOfCorrelations4sq}
\end{table}

\begin{table}[h]
 \mytablesize
 \tabcolsep 3pt
 \centering
 \begin{tabular}{c|ccc}
                     & $|r_s| \in [0.05,0.2)$   & $|r_s| \in [0.2,0.4)$  & $|r_s| \in [0.4,0.6)$\\
 \hline
 Greater London      & 4                     & 2                    & 0 \\
 Greater Manchester  & 1                     & 4                    & 0 \\
 West Midlands       & 1                     & 9                    & 0 \\
 \end{tabular}
 \caption{Number of OSM POI categories (positively or negatively) correlated with IMD (all results shown are statistically significant, $p$ $<$ 0.05)}
 \label{tab:numberOfCorrelationsOSM}
\end{table}

\begin{table}[h]
 \mytablesize
 \tabcolsep=1mm
 \centering
 \begin{tabular}{c|rrr}
                & \multicolumn{1}{|c}{\em Greater} & \multicolumn{1}{c}{\em Greater} &  \multicolumn{1}{c}{\em West}\\
                & \multicolumn{1}{|c}{\em London} & \multicolumn{1}{c}{\em Manchester} & \multicolumn{1}{c}{\em  Midlands} \\
 \hline
 Top            & Caribbean (0.37)         & Bus Station (0.32)    & Car Wash (0.38) \\
 positively     & African (0.32)           & Residential (0.27)    & Temple (0.34) \\
 correlated     & Fried Chicken (0.31)     & Student Centre (0.24) & Desserts (0.32) \\
    \hline
 Top            & Indian (-0.27)           & Italian (-0.36)       & Golf Course (-0.42) \\
 negatively     & Italian (-0.26)          & Golf Course (-0.28)   & Salon Barbershop (-0.35) \\
 correlated     & Golf Course (-0.24)      & Gastropub (-0.26)     & Farm (-0.31) \\
 \end{tabular}
 \caption{Foursquare categories most (positively and negatively) correlated with IMD. In parentheses, the Spearman correlation values (at statistical significance $p$ $<$ 0.05) computed with the Clifford {\em et al.} method are shown.}
 \label{tab:correlationsFoursquare}
\end{table}

\begin{table}[!h]
 \mytablesize
 \tabcolsep=1mm
 \centering
 \begin{tabular}{c|rrr}
                & \multicolumn{1}{|c}{\em Greater} & \multicolumn{1}{c}{\em Greater} &  \multicolumn{1}{c}{\em West}\\
                & \multicolumn{1}{|c}{\em London} & \multicolumn{1}{c}{\em Manchester} & \multicolumn{1}{c}{\em  Midlands} \\
 \hline
 Top            & traffic signals (0.29)  & traffic signals (0.25) & tram stop (0.31) \\
 positively     & crossing (0.25)         & taxi (0.24)            & billboard (0.29) \\
 correlated     & community center (0.18) &                        & artwork (0.29) \\
 \hline
 Top            & parking (-0.14)         & post box (-0.26)       & parking (-0.30) \\
 negatively     & garden center (-0.10)   & kindergarten (-0.22)   &   \\
 correlated     &                         & restaurant (-0.18)     &   \\
 \end{tabular}
 \caption{OSM categories most (positively and negatively) correlated with IMD. In parentheses, the Spearman correlation values  (at statistical significance $p$ $<$ 0.05)  computed with the Clifford {\em et al.} method are shown.}
 \label{tab:correlationsOSM}
\end{table}

\begin{table}[!h]
   \centering
   \mytablesize
   \begin{tabular}{c|c|cccc}
      \em Dataset & \em Urban area (\# categories) & \em 1st Qu. & \em Median & \em 3rd Qu. & \em Max. \\  
      \hline
                         & Greater London (35)                    & 0.04  & 0.14  & 0.30  & 0.40 \\
      Foursquare         & Greater Manchester (54)                & 0.02  & 0.03  & 0.05  & 0.09 \\
                         & West Midlands (50)                     & 0.06  & 0.10  & 0.14  & 0.20 \\
      \hline
                         & Greater London (6)                     & 0.32  & 0.32  & 0.63  & 0.63 \\
      OSM                & Greater Manchester (5)                 & 0.20  & 0.21  & 0.31  & 0.41 \\
                         & West Midlands (10)                     & 0.07  & 0.07  & 0.14  & 0.14 \\ 
  \end{tabular}
  \caption{$q$-values per quartiles of POI categories, computed using the False Discovery Rate technique.}
  \label{tab:falsepositives}
\end{table}

To quantify the expected proportion of false discoveries among the previously identified POI categories, we then followed our proposed methodology and applied the FDR control technique. More precisely, we first ranked our variables by their $p$-values (from the lowest to the highest), and then computed $q$-values on the ranked list. Table~\ref{tab:falsepositives} shows summary results: for each dataset (Foursquare and OSM), and for each city under study (Greater London, Greater Manchester and West Midlands), we report the computed $q$-values for each quartile of the (ranked) variables. Let us consider Foursquare-related results first: the expected proportion of false positive correlations is less than 9\% across all  POI categories considered significant for Greater Manchester, and at most 20\% for West Midlands. For Greater London, the expected proportion is at most 14\% for half of the discovered variables (those with lowest $p$), though it increases to 40\% when looking at the whole set. Results for OSM-related variables are less promising instead: with the exception of West Midlands, where the  $q$ values are low over the entire set of variables, for Greater London up to 63\% of the findings are expected to be false discoveries, and up to 41\% for Greater Manchester. For the purpose of the present study, the Foursquare dataset would thus appear more suited; indeed, the majority of significant POI categories, and the themes derived from them (which we are going to present next), come from Foursquare.

\subsection{Building Classifiers of Deprivation}

We now test whether we can exploit land use data extracted from user generated content to build accurate classifiers of urban deprivation. As we pointed out in the Method section, there are two possible ways to carry out this task: conducting household surveys on a small subset of city neighborhoods and estimating deprivation for the remaining ones; or conducting citywide surveys in one year and estimate deprivation for the subsequent ones. We evaluate the former approach next (we cannot test the latter, as we do not have UGC for different time steps). We proceeded as follows: we selected Greater London as case study (similar results were obtained for Greater Manchester and West Midlands), and divided IMD values into ten deciles, adhering to the methodology applied in the official IMD document~\cite{mclennan2011english}. We then randomly split our data in 25\% train and 75\% of  test, thus obtaining   156 neighborhoods for the train test and 469 neighborhoods for the test set respectively. We then built a variety of classifiers that take in input the {\em Offering Advantage} values of the Foursquare/OSM POI categories that show  statistically significant (positive or negative) correlations with IMD values in the training set. Next, we present results obtained using a Naive Bayes classifier -- the classifier that, among the tested ones (i.e., Decision Tree j48, Logistic Regression), showed the best performance.

Classification accuracy results are shown in Table~\ref{tab:ClassificationAccuracy10bins}. We note that the highest Precision and Recall values are obtained for classes $a$ and $j$, which represent the 10\% least deprived and the 10\% most deprived neighborhoods. This suggests that our method is most suitable to identify and monitor problematic areas, while performing less well for middle cases. Also, at first glance, Precision and Recall are not particularly high in absolute values. However, in almost all cases, the predicted class only differs of a few positions (two or three)  from the actual one, as evidenced by the Confusion Matrix reported in Table~\ref{tab:confusionMatrix10bins}.

\begin{table}[!h]
	\mytablesize
 	\centering
 	\begin{tabular}{cccl}
              \em Precision & \em Recall & \em F-Measure & \em Class\\
            \hline
               0.320         & 0.356      & 0.337         & $a$: 10\% more deprived\\
               0.189         & 0.227      & 0.206         & $b$: from 10\% to 20\% more deprived\\
               0.200         & 0.137      & 0.163         & $c$: from 20\% to 30\% more deprived\\
               0.135         & 0.102      & 0.116         & $d$: from 30\% to 40\% more deprived\\
               0.125         & 0.067      & 0.087         & $e$: from 40\% to 50\% more deprived\\
               0.143         & 0.080      & 0.103         & $f$: from 50\% to 60\% more deprived\\
               0.165         & 0.357      & 0.226         & $g$: from 60\% to 70\% more deprived\\
               0.194         & 0.255      & 0.220         & $h$: from 70\% to 80\% more deprived\\
               0.214         & 0.115      & 0.150         & $i$: from 80\% to 90\% more deprived\\
               0.262         & 0.364      & 0.305         & $j$: from 90\% to 100\% more deprived\\
 	\end{tabular}
 	\caption{Classification accuracy of urban deprivation for Greater London. IMD is subdivided in 10 bins.}
 	\label{tab:ClassificationAccuracy10bins}
\end{table}

\begin{table}[!h]
	\tabcolsep 3.85pt
	\mytablesize
 	\centering
 	\begin{tabular}{cccccccccc|l}
		 $a$ & $b$ & $c$ & $d$ & $e$ & $f$ & $g$ & $h$ & $i$ & $j$ & $\leftarrow$ classified as\\
		 \hline
         16 & 11 & 5 & 3 & 3 & 1 & 2 & 1 & 0 & 3 &$a$: 10\% more deprived\\
         10 & 10 & 4 & 4 & 4 & 0 & 7 & 3 & 1 & 1 &$b$: from 10\% to 20\% more deprived\\
         4 & 9 & 7 & 7 & 3 & 4 & 7 & 4 & 3 & 3 &$c$: from 20\% to 30\% more deprived\\
         7 & 4 & 4 & 5 & 3 & 3 & 17 & 2 & 2 & 2 &$d$: from 30\% to 40\% more deprived\\
         2 & 6 & 5 & 2 & 3 & 2 & 11 & 6 & 3 & 5 &$e$: from 40\% to 50\% more deprived\\
         4 & 4 & 6 & 2 & 1 & 4 & 12 & 4 & 4 & 9 &$f$: from 50\% to 60\% more deprived\\
         0 & 5 & 0 & 3 & 2 & 4 & 15 & 6 & 1 & 6 &$g$: from 60\% to 70\% more deprived\\
         2 & 3 & 1 & 5 & 1 & 3 & 11 & 12 & 3 & 6 &$h$: from 70\% to 80\% more deprived\\
         4 & 1 & 3 & 2 & 3 & 5 & 4 & 14 & 6 & 10 &$i$: from 80\% to 90\% more deprived\\
         1 & 0 & 0 & 4 & 1 & 2 & 5 & 10 & 5 & 16 &$j$: from 90\% to 100\% more deprived\\

 	\end{tabular}
 	\caption{Confusion Matrix associated with our classification model.}
 	\label{tab:confusionMatrix10bins}
\end{table}

At this point, one may wonder how our proposed method performs compared to both state-of-the-art approaches (e.g., \cite{eagle10,mao13,smith13a}) and simpler benchmarks derived from UGC datasets. To answer the former, we compare our classification results with those reported in \cite{smith13a}, where public transit data was used to estimate deprivation for Greater London, using the same ground truth data (IMD published in  2011) we rely upon. To ease comparison, since the work in  \cite{smith13a} divided the IMD distribution in two bins only (i.e., below and above the median value), we re-computed our classification using these two bins separated by the median value as output. To address the latter, we built a simple benchmark that estimates deprivation of a London ward by means of a Naive Bayes classifier that takes in input the number of Foursquare check-ins, the number of Foursquare POIs, and the number of OSM POIs present in that area; the intuition behind this benchmark is that, the higher the number of check-ins and POIs in a ward, the less deprived the ward is. 

Table~\ref{tab:ClassificationAccuracy2bins} shows results for our model, together with the  performance gain over the basic benchmark. Our model reaches a Precision between $0.763$ (for above-median deprivation) and $0.713$ (for below-median deprivation); the best performing classifier presented in \cite{smith13a} achieved an overall Precision of $0.805$; however, note that such result only holds for 10\% of the wards in London (where a tube station is present), while our results cover the whole of Greater London. By taking these two observations together, we argue that the performance of our classifier is indeed comparable to state-of-the-art approaches that require access to datasets that are not publicly available.  As for the performance of our model compared  to the simpler benchmark, we observe  significant improvements (shown in parentheses  in Table~\ref{tab:ClassificationAccuracy2bins}) for both Precision and Recall in both classification classes. This result demonstrates the suitability of the {\em Offering Advantage} metric over simple counts of check-ins and POIs.

\begin{table}[h]
	\tabcolsep 3.5pt
	\mytablesize	
 	\centering
 	\begin{tabular}{cccl}
        	\em Precision  & \em Recall    & \em F-Measure  & \multicolumn{1}{c}{\em Class} \\
		\hline
        	0.763 (+41\%)  & 0.692 (+17\%) & 0.726 (+28\%)  & 50\% more deprived \\
        	0.713  (+37\%) & 0.780 (+39\%) & 0.745 (+38\%)  & 50\% less deprived \\
 	\end{tabular}
 	\caption{Classification accuracy of urban deprivation for Greater London. IMD is subdivided in 2 bins. In parenthesis, the percentage difference w.r.t. the results of a basic benchmark is shown.}
 	\label{tab:ClassificationAccuracy2bins}
\end{table}

\subsection{Deriving Themes}

The previous results show that our method has competitive classification accuracy. To interpret those results, we conduct a thematic analysis \cite{bc06} of the POI categories identified through correlation analysis, and group them together in coherent themes. The analysis consisted of three iterations, separately conducted by an urban designer and a computer scientist. In the first iteration, the whole set of POI categories was scanned and initial codes were generated; this was followed by merging  semantically-related codes into broader themes using relevant urban studies as guidance; finally, identified themes were refined and named. In the end, a total of eight common themes were identified:  (derived from Foursquare) {\em health harmful food}, {\em faith}, {\em non-local cuisines}, {\em beauty \& aesthetics}, {\em sports}, {\em open spaces}, and {\em bus service}; (derived from OSM) {\em road system elements}. Table~\ref{tab:commonCorrelations} shows the Foursquare and OSM categories related to each theme, along with the Spearman correlation values for each category within these theme, computed through the Clifford {\em et al.} method~\cite{clifford89} between their {\em Offering Advantage} and IMD. Note that these correlation values are only valid for the three cities under study; as these three cities belong to the same country, it is not surprising that values for the same POI category are similar across them. However, if we were to apply this method to other cities in the world, the same POI category could bear opposite correlation with deprivation. While correlation findings are expected to differ, the same method could still be applied to other urban contexts, as long as UGC is available. We next briefly elaborate on each of the derived themes; in most cases, to gain confidence in their validity, we mention similar results in the literature.


\begin{table}
\tabcolsep 4pt
 \mytablesize
 \centering
 \begin{tabular}{l|l|lll}
 {\em Themes}   & {\em Category}     & \multicolumn{1}{|c}{\em Greater} & \multicolumn{1}{c}{\em Greater}    & \multicolumn{1}{c}{\em West} \\
                &                    & \multicolumn{1}{|c}{\em London}  & \multicolumn{1}{c}{\em Manchester} & \multicolumn{1}{c}{\em Midlands} \\
 \hline
 \multicolumn{5}{c}{\em Foursquare}\\
 \hline
 Health harmful & Fried Chicken       & 0.31          & 0.15              & 0.19     \\
 food           & Fast Food           &               & 0.22              & 0.31     \\
 				& Wings               & 0.11          &                   &          \\
 
 \hline 
 Faith          & Mosque              & 0.27          & 0.22              &          \\
                & Church              & -0.18         & -0.15             &          \\
                
 \hline 
 Non-local         & African             & 0.32          &                   & 0.25     \\
 cuisines       & Caribbean           & 0.37          &                   & 0.21     \\
                & Asian 	          &               &                   & 0.23     \\
                & Italian 	          & -0.26         & -0.36             & -0.25    \\
                & Indian 	          & -0.27         & -0.17             &          \\
                & Spanish 	          &               & -0.20             &          \\
                & Chinese 	          &               & -0.22             &          \\
                
 \hline
 Beauty \&      & Dentist's Office    & -0.22         & -0.21             & -0.15    \\
 aesthetics     & Nail Salon          &               & -0.17             & -0.19    \\
                & Salon Barbershop    & -0.15         &                   & -0.35    \\
                
 \hline                
 Sports         & Golf Course         & -0.24          & -0.28            & -0.42    \\
                & Cricket             & -0.13          &                  & -0.23    \\
                & Tennis Court        &                & -0.23            &          \\
                
 \hline                
 Open spaces    & Other Outdoors      & -0.15          & -0.24            & -0.25    \\
                & Lake                & -0.12          & -0.13            &          \\
                & Campground          &                & -0.22            & -0.23    \\
                & Field               & -0.15          & -0.23            &          \\
                & Playground          &                & -0.22            &          \\
                & Trail               &                & -0.21            &          \\
                & Outdoors and Recreation  &           & -0.14            &          \\      
               
 \hline                
 Bus service    & Bus                 & 0.15           &                  & 0.23     \\
                & Bus Station         & 0.28           & 0.32             &          \\
                & Bus Stop            & 0.18           &                  &          \\
                
 \hline
 \multicolumn{5}{c}{\em OSM}\\
 \hline
 Road system    & traffic signals     & 0.29           & 0.25             &          \\
 elements       & crossing            & 0.25           &                  &          \\
                & mini roundabout     &                &                  & 0.24     \\                                          
 \end{tabular}
 \caption{Spearman correlation values $r_s$ between Foursquare and OSM categories considered by the thematic analysis and IMD (all  results shown are statistically significant, $p$ $<$ 0.05).}
 \label{tab:commonCorrelations}
\end{table}

\paragraph{Health harmful food} We created this theme to include all the Foursquare venues that are related with restaurants selling unhealthy food. These are Fried Chicken, Fast Food and Wings. Those venues are positively associated with neighborhoods with IMD scores above the median. This finding is consistent with some studies in preventive medicine: using qualitative investigations, MacDonald {\em et al.} found that the higher the density of chain fast-food restaurants, the higher the neighborhood deprivation for England and Scotland~\cite{cummins05}. Other studies have been carried out in New Zealand \cite{pearce2007neighborhood} and in the USA \cite{block2004fast} and found similar results, thus suggesting the same correlation sign for this theme could be found in cities within these other countries too.

\paragraph{Faith} This theme includes two Foursquare venues: Mosque and Church. However, the two bear opposite correlation with deprivation: mosques tend to have higher concentration in areas with IMD above the median, while churches are more concentrated in areas with IMD below the median. Previous research has shown that there is a link between high concentrations of Muslim residents in London wards and below the median IMD values \cite{brimicombe2007ethnicity}. This seems to be consistent with part of our finding; however it is also true that Muslims might not live where their  places of worship are located. To ascertain this missing point, we studied the relationship between the percentage of Muslims living in a certain area (relative to the total number of religious people in that area) and the {\em Offering Advantage} for the Foursquare category Mosque. We did so by extracting information from the Census Data 2011 for Greater London at the level of ward.\footnote{\url{http://www.nomisweb.co.uk/census/2011/ks209ew}} We indeed found a positive correlation between the presence of mosques in a neighborhood and percentages of Muslim residents in it ($r_s=0.40$, $p$ $<$ 0.01). This seems to be congruent with the hypothesis that neighborhoods with Muslim predominance, which are generally associated with above-median IMD values in Greater London \cite{brimicombe2007ethnicity}, have a higher-than-normal number of mosques. 

\paragraph{Non-local cuisines} We created this theme to include all Foursquare venues that are related to restaurants but exclude those covering local cuisine (i.e., Pub, Fish \& Chips Shop, English Restaurant), as the latter did not bear strong correlation with IMD. Within this broad theme, we identified two sub-themes: one comprising cuisines that, in the cities under consideration, had positive correlation with deprivation (e.g., African, Asian and Caribbean), and one comprising cuisines that, once again for the three cities under consideration, had negative correlation with deprivation (e.g., Italian, Chinese, Spanish, Indian). 


\paragraph{Beauty \& aesthetics} This theme comprises three Foursquare venues: Dentist, Nail Salon and Salon Barbershop.  All these categories are negatively correlated with IMD, suggesting that beauty and aesthetics facilities concentrate in neighborhoods with IMD below the median. We found a reference to this finding for the category Dentist's Office. Previous studies have, in fact, demonstrated that high socioeconomic status is significantly associated with good oral health \cite{locker2000deprivation}; our results seem to be congruent with those findings. However, this is an example of a finding that may not generalise to other geographic contexts: for example, previous research has found a link between the prevalence of beauty salons in areas of the USA  and their socio-economic deprivation~\cite{small2006}. Note that, although the finding itself might not generalise, the methodology we propose to discover whether this is indeed the case remains the very same.

\paragraph{Sports} This theme includes all the Foursquare categories related with physical activity facilities. These are Golf Course, Cricket and Tennis Court. These venues are negatively correlated with deprivation, suggesting that sport facilities tend to be concentrated in areas with low IMD scores. Previous qualitative studies are consistent with this finding: researchers found that golf courses in Australia \cite{giles02}, fitness centers and dance facilities in the USA \cite{powell06} tend to be more commonly available in wealthier areas.

\paragraph{Open spaces} We create this theme to include all the Foursquare facilities related to public open spaces. These are Lake, Outdoors and Recreation, Playground, Campground, Trail, Field and Other Outdoors. All these categories are negatively correlated with deprivation. Previous studies are congruent with these results and found similar outcomes for the Netherlands \cite{luttik2000value}, in Howard County,  USA \cite{geoghegan2002value} and for Portland, USA \cite{bolitzer2000impact}.

\paragraph{Bus service} This theme includes the Foursquare categories for Bus, Bus Stop and Bus Station.  These venues are positively correlated with deprivation; their concentration in an area is thus signal of higher than average deprivation. This finding is consistent with a recent report issued by Transport for London that shows that, for Greater London, the share of bus trips increases with the decrease of household incomes \cite{tfl2011}. 

\paragraph{Road system elements} This theme includes all the OSM amenity types related with the traffic management system. These are traffic signals, crossings (i.e., elements which guarantee safe pedestrian crossing) and mini roundabouts. All of these OSM amenities are positively correlated with deprivation and therefore tend to be concentrated in areas with IMD scores above the median. Higher-than-normal  presence of these elements may be associated with road infrastructures (e.g.,  junctions, highways, main roads), which might make a neighborhood less attractive to live in.

%% file: implications.tex
In the previous sections, we have proposed and evaluated a new method that mines urban data obtained from easily accessible UGC datasets (namely, Foursquare and OpenStreetMap) to extract features that  correlate with metrics of socioeconomic deprivation at the level of cities' neighborhoods. We have shown that we can use these features to build accurate classifiers of deprivation and that the corresponding results are in line with previous findings. 

\paragraph{Implications} This work has both practical and theoretical implications. From a practical standpoint, the suggested method affords the ability to build `neighborhood profiling' tools that  different stakeholders can use for different purposes: for example, residents may use them to decide where to buy a property or rent a flat; visitors may consult them to decide in which hotel or rent-house to stay; and city planners and administrators may use them to analyze and compare what makes a well-off \emph{vs.} a deprived neighborhood in their cities. From a theoretical standpoint, our method can be used by urban designers and social science researchers to advance knowledge in their fields: for example,  to understand the relationship between the built environment and deprivation, as it varies across different cities and cultures; and to analyze the relationships between the built environment and deprivation as it varies in response to different processes such as urbanization and gentrification. Note that, when analysing different geographic contexts, findings elicited with our methodology may well differ; however, the methodology itself is generally applicable, and indeed can be applied in different geographic settings to discover these variations.

\paragraph{Limitations} When selecting to apply our proposed method, one needs to take into consideration the following limitations. First, both Foursquare and OSM datasets have geographic and social biases. The two datasets, in fact, do not have a uniform coverage of urban features across space; rather, coverage is  concentrated in city centers, thus affording us only a partial picture of what elements are present in areas further away~\cite{Mashhadi2013cscw}. In terms of social bias, both Foursquare and OSM users belong to a rather specific demographic group (i.e., young, educated, wealthy); one may thus question how representative the data they produce is of what indeed exists in the physical space.  When applying our methodology, one should first check for geographic and social biases, for example using the method proposed in \cite{cscw15gio};  for cities where such effects are large,  our methodology is more likely to return invalid results. 

The second limitation which ought to be acknowledged concerns the multiple comparison problem. Our method requires the simultaneous test of multiple variables, and this might increase the chance of finding false positive correlations.  Statistical control techniques like FDR should be applied, before deciding whether risks of invalid findings are low.  In the case studies presented in this work, we applied  FDR to estimate such risk; results appear robust when using Foursquare data, especially for the cities of Greater Manchester and West Midlands.

The third limitation has to do with the coarse-grained classification granularity. We have demonstrated clear performance advantage of our method over a plausible baseline and over a state-of-the-art method that classifies urban deprivation in a binary fashion. In the future, to perform finer-grained classifications, researchers might use  multi-modal machine learning approaches to combine features derived from multiple datasets. For example, they could  combine social media data with image data coming from Google Street Views, to obtain information about the aesthetic capital afforded by many neighborhoods around the world~\cite{hwang14,quercia14aesthetic}.

Finally, for moderate levels of urban deprivation, classification accuracy should be improved, since
our classifier performs really well only at the two extremes of the distribution. One way of doing so is to explore metrics other than Offering Advantage. There might be metrics that capture different notions of  a neighborhood's potential offering  (e.g., what types of POIs are within walking distance) and that express those notions with alternative mathematical formulations (e.g., graph-based ones~\cite{hillier1984social}).


\paragraph{Future work} Our future work spans two main directions: on one hand, we aim to expand the method so to capture other urban features (e.g., walkability) that pertain the physical layout of neighborhoods. OSM data is particularly well-suited to extract such features, as it comprises not just POI data (as is the case for Foursquare), but also road network information. Examples of features we might computationally capture include average block size and road  density, that urban studies have previously linked to indicators of neighborhood well-being \cite{ref1,ref2}. On the other hand, we aim to quantify the applicability of our method to measure deprivation in different geographic contexts, especially in developing countries where accurate and up-to-date indexes of deprivation like IMD are very difficult to build. We note that, in these contexts, UGC data as obtained by social media services like Foursquare is probably of very little value (too sparse); however, OSM penetration has been shown to be high also in developing countries~\cite{quattrone2014mind}, thus suggesting our method could be used in these more challenging contexts too, especially once we have expanded our set of features to comprise those that we can extract from the road network.

%% file: cscw15_venerandi.bbl
\begin{thebibliography}{10}

\bibitem{ref2}
Alberti, M.
\newblock The effects of urban patterns on ecosystem function.
\newblock {\em International Regional Science Review 28}, 2 (2005), 168--192.

\bibitem{block2004fast}
Block, J.~P., Scribner, R.~A., and DeSalvo, K.~B.
\newblock Fast food, race/ethnicity, and income: a geographic analysis.
\newblock {\em American journal of preventive medicine 27}, 3 (2004), 211--217.

\bibitem{bolitzer2000impact}
Bolitzer, B., and Netusil, N.~R.
\newblock {The impact of open spaces on property values in Portland, Oregon}.
\newblock {\em Journal of environmental management 59}, 3 (2000), 185--193.

\bibitem{bc06}
Braun, V., and Clarke, V.
\newblock Using thematic analysis in psychology.
\newblock {\em Qualitative Research in Psychology 3}, 2 (2006), 77--101.

\bibitem{brimicombe2007ethnicity}
Brimicombe, A.~J.
\newblock {Ethnicity, religion, and residential segregation in London: evidence
  from a computational typology of minority communities}.
\newblock {\em ENVIRONMENT AND PLANNING B PLANNING AND DESIGN 34}, 5 (2007),
  884.

\bibitem{clifford89}
Clifford, P., Richardson, S., and Hemon, D.
\newblock {Assessing the significance of the correlation between two spatial
  processes}.
\newblock {\em Biometrics 45}, 1 (1989), 123--134.

\bibitem{cranshaw12}
Cranshaw, J., Schwartz, R., Hong, J.~I., and Sadeh, N.
\newblock {The Livehoods Project: utilizing social media to understand the
  dynamics of a city}.
\newblock In {\em Proc. of ICWSM}, AAAI (2012).

\bibitem{cummins05}
Cummins, S. C.~J., McKay, L., Witten, K., and MacIntyre, S.
\newblock {McDonalds restaurants and neighborhood deprivation in Scotland and
  England}.
\newblock {\em American Journal of Preventive Medicine 29}, 4 (2005), 308--310.

\bibitem{pdun11}
{Department of Economic and Social Affairs}.
\newblock World urbanization prospects, the 2011 revision: highlights.
\newblock Tech. rep., Population Division United Nations, 2011.

\bibitem{eagle10}
Eagle, N., Macy, M., and Claxton, R.
\newblock {Network diversity and economic development}.
\newblock {\em Science 328\/} (2010), 1029--1031.

\bibitem{elvidge97}
Elvidge, C.~D., Baugh, K.~E., Kihn, E.~A., Kroehl, H.~W., and Davis, E.~R.
\newblock {Mapping city lights with nighttime data from the DMSP Operational
  Linescan System}.
\newblock {\em Photogrammetric engineering and Remote Sensing 63}, 6 (1997),
  727--734.

\bibitem{elvidge01}
Elvidge, C.~D., Imhoff, M.~L., Baugh, K.~E., Hobson, V.~R., Nelsonc, I.,
  Safran, J., Dietz, J.~B., and Tuttle, B.~T.
\newblock {Night-time lights of the world: 1994-1995}.
\newblock {\em Photogrammetry and Remote Sensing 56}, 2 (2001), 81--99.

\bibitem{geoghegan2002value}
Geoghegan, J.
\newblock The value of open spaces in residential land use.
\newblock {\em Land use policy 19}, 1 (2002), 91--98.

\bibitem{giles02}
Giles-Corti, B., and Donovan, R.
\newblock {Socioeconomic status differences in recreational physical activity
  levels and real and perceived access to a supportive physical environment}.
\newblock {\em Preventive Medicine}, 35 (2002), 601--611.

\bibitem{grenyer06}
Grenyer, R., Orme, C. D.~L., Jackson, S.~F., Thomas, G.~H., Davies, R.~G.,
  Davies, T.~J., Jones, K.~E., Olson, V.~A., Ridgely, R.~S., Rasmussen, P.~C.,
  Ding, T., Bennett, P.~M., Blackburn, T.~M., Gaston, K.~J., Gittleman, J.~L.,
  and Owens, I. P.~F.
\newblock {Global distribution and conservation of rare and threatened
  vertebrates}.
\newblock {\em Nature 444\/} (2006), 93--96.

\bibitem{griffith11}
Griffith, D.~A., and Paelinck, J.~H.
\newblock {\em Non-standard spatial statistics and spatial econometrics}.
\newblock Springer, 2011.

\bibitem{hidalgo2007product}
Hidalgo, C.~A., Klinger, B., Barab{\'a}si, A.~L., and Hausmann, R.
\newblock The product space conditions the development of nations.
\newblock {\em Science 317}, 5837 (2007), 482--487.

\bibitem{hillier1984social}
Hillier, B., and Hanson, J.
\newblock {\em The social logic of space}.
\newblock {Cambridge University Press}, 1984.

\bibitem{hwang14}
Hwang, J., and Sampson, R.~J.
\newblock {Divergent Pathways of Gentrification: Racial Inequality and the
  Social Order of Renewal in Chicago Neighborhoods}.
\newblock {\em {American Sociological Review}\/} (2014).

\bibitem{Lacy2011}
Lacy, S.
\newblock {Foursquare closes \$50M at a \$600M valuation}.
\newblock Tech. rep., TechCrunch, 2011.

\bibitem{legendre93}
Legendre, P.
\newblock {Spatial autocorrelation: trouble or new paradigm?}
\newblock {\em Ecology 74}, 6 (1993), 1659--1673.

\bibitem{lindqvist2011m}
Lindqvist, J., Cranshaw, J., Wiese, J., Hong, J., and Zimmerman, J.
\newblock {I'm the mayor of my house: examining why people use Foursquare - a
  social-driven location sharing application}.
\newblock In {\em Proc. of CHI}, ACM (2011), 2409--2418.

\bibitem{locker2000deprivation}
Locker, D.
\newblock Deprivation and oral health: a review.
\newblock {\em Community dentistry and oral epidemiology 28}, 3 (2000),
  161--169.

\bibitem{luttik2000value}
Luttik, J.
\newblock {The value of trees, water and open space as reflected by house
  prices in the Netherlands}.
\newblock {\em Landscape and Urban Planning 48}, 3 (2000), 161--167.

\bibitem{mao13}
Mao, H., Shuai, X., Ahn, Y.~Y., and Bollen, J.
\newblock {Mobile communications reveal the regional economy in Cote d'Ivoire}.
\newblock In {\em Proc. of NetMob} (2013).

\bibitem{Mashhadi2013cscw}
Mashhadi, A., Quattrone, G., and Capra, L.
\newblock {Putting Ubiquitous Crowd-sourcing into Context}.
\newblock In {\em Proc. of CSCW}, ACM (2013), 611--622.

\bibitem{mclennan2011english}
McLennan, D., Barnes, H., Noble, M., Davies, J., Garratt, E., and Dibben, C.
\newblock The english indices of deprivation 2010.
\newblock {\em London: Department for Communities and Local Government\/}
  (2011).

\bibitem{unhsp08}
Moreno, E.~L., Bazoglu, N., Mboup, G., and Warah, R.
\newblock State of the world's cities 2008/2009 - harmonious cities.
\newblock Tech. rep., UN-HABITAT, 2008.

\bibitem{noor08}
Noor, A.~M., Alegana, V.~A., Gething, P.~W., Tatem, A.~J., and Snow, R.~W.
\newblock {Using remotely sensed night-time light as a proxy for poverty in
  Africa}.
\newblock {\em Population Health Metrics 6}, 5 (2008), 81--99.

\bibitem{odi08}
{Overseas Development Institute}.
\newblock Briefing paper 44: Opportunity and exploitation in urban labour
  markets.
\newblock Tech. rep., 2008.

\bibitem{pearce2007neighborhood}
Pearce, J., Blakely, T., Witten, K., and Bartie, P.
\newblock Neighborhood deprivation and access to fast-food retailing: a
  national study.
\newblock {\em American journal of preventive medicine 32}, 5 (2007), 375--382.

\bibitem{powell06}
Powell, L., Slater, S., Chaloupka, F., and Harper, D.
\newblock {Availability of physical activity-related facilities and
  neighborhood demographic and socioeconomic characteristics: a national
  study}.
\newblock {\em American Journal of Public Health 96}, 9 (2006), 1676--1680.

\bibitem{cscw15gio}
Quattrone, G., Capra, L., and Meo, P.~D.
\newblock {There’s No Such Thing as the Perfect Map: Quantifying Bias in
  Spatial Crowd-sourcing Datasets}.
\newblock In {\em {Proc. of CSCW}}, ACM (2015).

\bibitem{quattrone2014mind}
Quattrone, G., Mashhadi, A., and Capra, L.
\newblock Mind the map: the impact of culture and economic affluence on
  crowd-mapping behaviours.
\newblock In {\em Proc. of CSCW}, ACM (2014).

\bibitem{quercia12a}
Quercia, D., Ellis, J., Capra, L., and Crowcroft, J.
\newblock {Tracking Gross Community Happiness from Tweets}.
\newblock In {\em Proc. of CSCW}, ACM (2012).

\bibitem{quercia14aesthetic}
Quercia, D., Ohare, N., and Cramer, H.
\newblock {Aesthetic Capital: What Makes London Look Beautiful, Quiet, and
  Happy?}
\newblock In {\em {Proc. of CSCW}}, ACM (2014).

\bibitem{quercia2014mining}
Quercia, D., and Saez, D.
\newblock {Mining urban deprivation from Foursquare: implicit crowdsourcing of
  city land use}.
\newblock {\em Pervasive Computing, IEEE 13}, 2 (2014), 30--36.

\bibitem{quercia12b}
Quercia, D., S\'eaghdha, D.~O., and Crowcroft, J.
\newblock Talk of the city: our tweets, our community happiness.
\newblock In {\em Proc. of ICWSM}, AAAI (2012).

\bibitem{small2006}
Small, M.~L., and McDermott, M.
\newblock The presence of organizational resources in poor urban neighborhoods:
  An analysis of average and contextual effects.
\newblock {\em Social Forces 84}, 3 (2006), 1697--1724.

\bibitem{smith13a}
Smith, C., Mashhadi, A., and Capra, L.
\newblock Ubiquitous sensing for mapping poverty in developing countries.
\newblock In {\em Paper submitted to the Orange D4D Challenge (2013)} (2013).

\bibitem{smith13b}
Smith, C., Quercia, D., and Capra, L.
\newblock Finger on the pulse: identifying deprivation using transit flow
  analysis.
\newblock In {\em Proc. of CSCW}, ACM (2013), 683--692.

\bibitem{storey2003}
Storey, J.~D., and Tibshirani, R.
\newblock Statistical significance for genomewide studies.
\newblock In {\em Proc. of NAS} (2003), 9440--9445.

\bibitem{tobler70}
Tobler, W.~R.
\newblock {A computer movie simulating urban growth in the Detroit region}.
\newblock {\em Economic Geography 46\/} (1970), 234--240.

\bibitem{tfl2011}
{Transport for London (TFL)}.
\newblock {Travel in London, Supplementary Report: London Travel Demand Survey
  (LTDS)}.
\newblock Tech. rep., 2011.

\bibitem{ref1}
Tratalos, J., Fuller, R.~A., Warren, P.~H., Davies, R.~G., and Gaston, K.~J.
\newblock Urban form, biodiversity potential and ecosystem services.
\newblock {\em Landscape and Urban Planning 83}, 4 (2007), 308 -- 317.

\end{thebibliography}
